\documentclass{article}
\usepackage{spconf,amsmath,graphicx,amssymb,booktabs,bm,pifont,multirow}
\usepackage[hidelinks]{hyperref}
\newcommand*{\email}[1]{\href{mailto:#1}{\nolinkurl{#1}} } 

\title{Unsupervised Cross-Lingual Speech Emotion Recognition Using Pseudo Multilabel}
\name{Jin Li$^{1,2}$, Nan Yan$^{1,2}$, Lan Wang$^{1,2}$}
\address{$^1$CAS Key Laboratory of Human-Machine Intelligence-Synergy Systems, \\
	 Shenzhen Institute of Advanced Technology, Chinese Academy of Sciences, Shenzhen, China\\
	$^2$Guangdong-Hong Kong-Macao Joint Laboratory of Human-Machine Intelligence-Synergy Systems, \\
	 Shenzhen Institute of Advanced Technology, Chinese Academy of Sciences, Shenzhen, China \\
	\email{{li.jin, nan.yan, lan.wang}@siat.ac.cn}
}

\begin{document}
%\ninept
%
\maketitle
\begin{abstract}
Speech Emotion Recognition (SER) in a single language has achieved remarkable results through deep learning approaches in the last decade. However, cross-lingual SER remains a challenge in real-world applications due to a great difference between the source and target domain distributions. To address this issue, we propose an unsupervised cross-lingual Neural Network with Pseudo Multilabel (NNPM) that is trained to learn the emotion similarities between source domain features inside an external memory adjusted to identify emotion in cross-lingual databases. NNPM introduces a novel approach that leverages external memory to store source domain features and generates pseudo multilabel for each target domain data by computing the similarities between the external memory and the target domain features. We evaluate our approach on multiple different languages of speech emotion databases. Experimental results show our proposed approach significantly improves the weighted accuracy (WA) across multiple low-resource languages on Urdu, Skropus, ShEMO, and EMO-DB corpus. To facilitate further research, code is available at https://github.com/happyjin/NNPM
\end{abstract}
\section{Introduction}
\label{sec:intro}

Speech emotion recognition(SER) is the recognition of different human emotions from a given speech and is gaining increasing interest in the areas of human-computer interaction, computational neuroscience, cognitive psychology, intelligent tutor for children, and medical healthcare \cite{el2011survey, cowie2001emotion, albu2015neural}. Advancements in machine learning methods in recent years have allowed SER systems to exhibit excellent performance when training and testing data from corpora of the same language. However, the development of more robust SER systems for the cross-lingual scenarios remains an open problem due to the domain mismatch problem.

To date, several supervised methods have been proposed to reduce the domain mismatch problem for cross-lingual SER systems. One approach involves training with a combination of diverse corpora, which is able to reduce the factors from varying conditions such as acoustic environment and improves the model performance \cite{schuller2010cross}. Class-wise adversarial domain adaptation is another method to solve this problem by reducing the domain shift for all classes between different corpora \cite{zhou2019transferable}. However, labels of some languages are costly to acquire, especially for low-resource languages, leading to a lack of available data for developing SER systems. This drawback limits the application of supervised methods.

\begin{figure*}[!h]
	\centering
	\includegraphics[width=15cm, height=8cm]{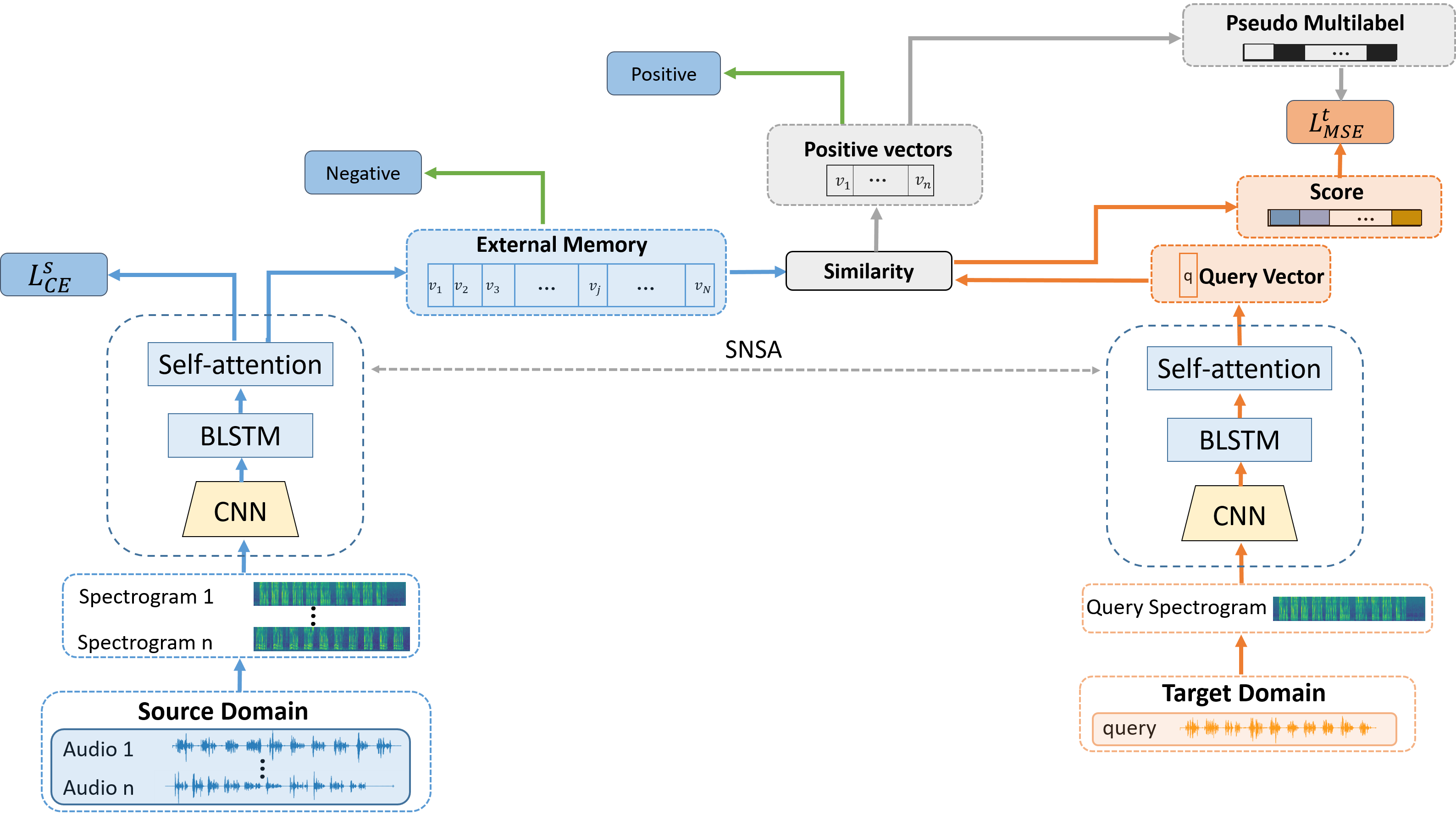}
	\caption{The pipeline of our proposed NNPM. We first compute spectrograms for both source domain audios and query audio from the current minibatch. Then both spectrograms are input into Siamese Network with Self Attention (SNSA) and the feature is extracted from this block. The source-domain features are dynamically stored in and read from the dynamic external memory. To learn the emotion similarities by computing the cosine distance between features inside dynamic external memory and the query vector. The query vector is a feature extract from the SNSA of the target domain. After that, the query vector from the target domain is assigned with pseudo multilabel based on the similarity scores. The model is optimized through loss from the source domain and the target domain using a hard negative sample mining strategy.}
	\label{fig:architecture}
\end{figure*}

Unsupervised methods that do not need data labels in the target domain provide an alternative solution for cross-lingual SER, especially for low-resource languages. Some cross-lingual SER systems based on unsupervised learning methods are introduced recently \cite{latif2019unsupervised, ahn2021cross, abdelwahab2018domain, kim2017towards}. A mainstream method is to utilize the adversarial-based method since it can learn the corpus-invariant representations using domain adversarial training. Generative Adversarial Network (GAN)-based method is proposed by unsupervised domain adaptation for multilingual SER and learns language invariant feature representations from source language features to target language features \cite{latif2019unsupervised}. However, GAN is hard to train and prone to suffer from convergence failure \cite{goodfellow2016nips}. Domain adversarial neural network (DANN) is also a method by generating a domain invariant feature representation that reduces the gap between features in the source and target domain \cite{abdelwahab2018domain}. However, the effectiveness of domain adversarial training is highly correlated with the distribution of two databases so that adversarial attacks and instabilities may occur during training if the data points are sharply different from each other \cite{zhang2019limitations}. Multi-task learning training methodology for unsupervised cross-lingual is another approach to improve the generalisability of the model by incorporating information on gender and naturalness \cite{kim2017towards}. An alternative approach is proposed by training to learn the emotion similarities from source domain samples through few-shot learning adapted to the target domain \cite{ahn2021cross}. However, samples in few-shot learning significantly depend on the choice of support set that might make the challenging to apply to practical settings. Moreover, because of the strong assumption of the support set sampled uniformly from a single distribution, it leads to the unstable number of samples selected for each class during the training process \cite{ochal2020comparison}. 

In an attempt to provide a solution to the challenges described above, we build a novel framework that models the unsupervised cross-lingual SER as a multilabel classification problem. This idea is inspired by ensemble learning \cite{opitz1999popular}, which averages the decisions from multiple related emotion features and decreases the effect from one specific emotion label to improve the model robustness. In addition, a dynamic external memory was also designed to store and update source domain features stably so that the number of samples from the source domain would be all utilized during training. As illustrated in Figure \ref{fig:architecture}, given two groups of audios from the source domain and the target domain, a model is first trained based on the source domain in a supervised way. Then the source domain features are saved in dynamic external memory and pseudo multilabel is assigned to target domain data by computing the similarities between features of the dynamic external memory and features of the target domain. Finally, the unsupervised training phase is optimized by combining the source domain cross entropy loss and loss from the target domain. Experimental results show the effectiveness of our NNPM in series of experiments.

The main contributions of this paper are: (1) a novel unsupervised cross-lingual SER framework with pseudo multilabel in the target domain; (2) a dynamic external memory design with memory update mechanism; (3) vastly exceed the baseline and other approaches for unweighted accuracy cross different languages.

The rest of this paper is organized as follows. The baseline model and the proposed method are described in Section 2. The experimental setup and results analysis are present in Section 3. Section 4 accomplishes the study with conclusions and directions.

\section{Methodology}

\subsection{Siamese Network with Self Attention (SNSA)}

Let $(X_s, Y_s):=\{(x_i^s,y_i^s)\}_{i=1}^n$ denotes the spectrogram $x^s \in X_s$ and its corresponding label $y^s \in Y_s$ in the source domain, where $n$ is the number of data in the source domain, and $X_t=\{x_i^t\}_{i=1}^m$ denotes the spectrogram of the target domain corpus without labeling, where $x_i^s \in \mathbb{R}^{L\times d}$ and $x_i^t \in \mathbb{R}^{L\times d}$ represent $L$ the temporal length of the spectrogram in the source and the target domain separately and $d$ is the dimension of a spectrogram feature vector. 

The SNSA is used to extract the features of the source domain spectrogram $X_s$ and the target domain spectrogram $X_t$. Our SNSA consists of two identical self-attention modules with shared weights for each other. For each self-attention module, we follow the previous work \cite{li2019improved} that mainly consists of 3 components, namely a 2-layer convolutional neural network (CNN), a 2-layer bidirectional LSTM (BLSTM) \cite{hochreiter1997long}, and a self-attention network. A temporal MaxPooling is applied to each convolution layer for dimensionality reduction and ReLU \cite{nair2010rectified} is applied as an activation function after each MaxPooling operation to enhance the model non-linearity. The BLSTM with forward and backward hidden states concatenation is represented by $H^{\text{blstm}}$. The self-attention $a$ is the weighted sum of the hidden states given $H^\mathrm{blstm}$ and is described by the following equations
\begin{equation}
	a = \mathrm{softmax}(W_2 \tanh (W_1 H^{\mathrm{blstm}}))
\end{equation}
\begin{equation}
	h^{\mathrm{attn}} = a H^{\mathrm{blstm}}
\end{equation} 
where $W_1$ and $W_2$ are the parameters to be optimized during training. $h^{\mathrm{attn}}$ can be either source domain feature $h^{\mathrm{attn}}_s$ or target domain feature $h^{\mathrm{attn}}_t$.

\subsection{Dynamic External Memory}

Dynamic external memory is proposed to store and load source domain features dynamically. Writing operation stores and updates the source domain features into the dynamic external memory. Specifically, it is updated by removing previous features and writing new features with a decay rate into the dynamic external memory. One crucial problem for the dynamic external memory update can be correlated with the stability-plasticity dilemma \cite{grossberg1987competitive}. On the one hand, dynamic external memory updates should be stable to avoid mismatching problems. Frequently changing the features in the dynamic external memory in every iteration decrease the training stability. On the other hand, the system also needs enough plasticity to effectively incorporate new features from the source domain into the dynamic external memory. In response to this issue, we purpose a weighted memory update strategy for the writing operation, which can be formulated as follows:
\begin{equation}
	\mathbf{M}^\mathrm{iter}[A] = || \beta \mathbf{M}^\mathrm{iter}[A] + (1-\beta)\mathbf{M}^{\mathrm{iter} - 1}[A] ||_2
\end{equation}
where $||\cdot||_2$ represents $\ell_2$ norm, $\mathbf{M} \in \mathbb{R}^{n\times d}$ is a dynamic external memory to save features and $\mathbf{M}[A_i]=h^\mathrm{attn}_i$, $A$ is an index set described in Equation \ref{eq:3}, $\beta$ is an updating rate that controls how fast the source domain features are written into the dynamic external memory. The reading operation loads features from the dynamic external memory and can be formulated as $\mathbf{M}^\mathrm{read} \gets \mathbf{M}^\mathrm{iter}[A]$.

\subsection{Pseudo-Labeling}

Since the target domain labels are not available in the proposed framework, pseudo multilabel is assigned for each data in the target domain. Pseudo-multi-labeling is a process of picking up multilabel for the target domain data. 

Each $x^t_i\in X_t$ is assigned with a pseudo label $\tilde{y}^t_i$. The assigning pseudo multilabel process can be expressed as follows
\begin{equation}
	\tilde{y}^t_i = Y_s[j] \quad \text{for } j \in A
	\label{eq:3}
\end{equation}
where $A$ is an index set of source labels and for each index satisfies the similarity scores between source domain features of the read dynamic external memory as well as a feature from the target domain is higher than the similarity score threshold $\gamma$. This process can be formulated by the inner product between the external memory and target domain feature $|| \langle\ \mathbf{M}^\mathrm{read}, h_j^{\mathrm{attn}} \rangle ||_2 \geq \gamma$, where $<\cdot,\cdot>$ is the inner product. With each unlabeled data in the target domain assigned pseudo label $\tilde{y}^t_i$, we can reformulate the target domain corpus as $(X_t,\tilde{Y}_t)=\{(x_i^t,\tilde{y}^t_i)\}_{i=1}^n$. With the pseudo multilabel $\tilde{y}^t_i$, data from the source domain and the target domain can be combined and train them together by the supervised loss function.

\subsection{Loss Functions}

The loss function of the proposed NNPM consists of two parts, one from the source domain, and one from the target domain.

The source domain loss function aims to bridge the gap between ground truth labels and predictions from source domain data under supervised learning. This function can be expressed by cross-entropy loss function as
\begin{equation}
	\mathcal{L}_\text{CE}^s = - \mathbb{E}_{(x^s_k,y^s_k)\sim (X_s, Y_s)} \sum_{k=1}^{n} y^s_k \log p(x^s_k)
\end{equation}
where $y^s_i$ is the ground-truth label of source domain data, $n$ is the batch size and $p(x^s_k)$ denotes the predicted classification probability after softmax function.

For target domain loss, the mean square loss (MSE) function is utilized as follows 
\begin{equation}
	\label{eq:mse}
	\mathcal{L}^t = \frac{1}{n} \sum_{k=1}^{n} || y^t_k - \tilde{y}_k ||^2_2
\end{equation}
where $y^t_k$ is the prediction by matrix multiplication between the dynamic external memory features $\mathbf{M}^\mathrm{read}$ and target domain features in the current batch. 

The dynamic external memory contains a large number of features that are not assigned with labels and are treated as negative samples after the pseudo-multi-labeling process. As a result, the target domain suffers from a sample imbalance problem between positive and negative samples. To moderate this problem, hard negative sample mining \cite{shrivastava2016training} is applied to focus more on hard negative samples than easy negative samples to make the model more robust and discriminative. This process can be formulated as 
\begin{equation}
	\label{eq:hard}
	H_{\mathrm{neg}} = \{N_i | N_i= \lambda \mathrm{sort} (N_{ \mathrm{neg}}) \}
\end{equation}
where $N_\mathrm{neg}$ is the index of negative sample score, $\mathrm{sort}(\cdot)$ represents sorted negative score from large to small and $\lambda$ is the hard negative sample ratio. With this hard negative sample mining process, Equation (\ref{eq:mse}) can be reformulated as the summation of two parts
\begin{equation}
	\mathcal{L}^t_\mathrm{hard} = \mathcal{L}^t_\mathrm{pos} + \mathcal{L}^t_\mathrm{neg}
\end{equation}
where $\mathcal{L}^t_\text{pos}$ is the MSE for positive samples and $\mathcal{L}^t_\text{neg}$ is MSE for the positive samples and $H_{\mathrm{neg}}$ hard negative samples.

Finally, the overall training loss of this work is given by:
\begin{equation}
	\mathcal{L} = \mathcal{L}_\mathrm{CE}^s + \mathcal{L}_\mathrm{hard}^t
\end{equation}

\section{Experiments}

\subsection{Databases}

\subsubsection{IEMOCAP}

The IEMOCAP \cite{busso2008iemocap} was an audiovisual database developed to investigate the connection between gestures, speech, and emotions and contained a total of five sessions recorded by 10 professional actors (5 males and 5 females). The database was segmented by speaker turn and 10,039 utterances were generated with 5,255 scripted turns and 4,787 improvised turns respectively. Four out of ten emotional categories provided by the corpus were used in the experiment. They are angry, happy, sad, and neutral, where the happy comprises the samples labeled as excited. 

\subsubsection{Urdu}

The Urdu language is an official language in Pakistan. The Urdu database was comprised of spontaneous emotional speech collected from Urdu TV talk show \cite{latif2018cross}. The database contained a total of 400 utterances from 38 speakers (27 males and 11 females) with four categorical emotional categories: angry, happy, sad, and, neutral. The corpus included emotional excerpts from spontaneous unscripted discussions among different speakers on the TV talk show. The data were split into training and test datasets using the Scikit-learn toolkit \cite{pedregosa2011scikit} with a ratio of 67\% for the training set and 33\% for the test set respectively.

\subsubsection{Estonian}
The Estonian emotional speech corpus (Ekropus) contained recordings of read speech sentences of four categorical emotions: anger, joy, sadness, and neutral \cite{vcermakhlt}. Out of the 173 sentences (1473 tokens) in total, 45.7\% was anger, 11.6\% was sadness, 1.7\% was joy (happiness), and 14.4\% was neutral. The dataset partition followed the same ratio that for the Urdu corpus. 

\subsubsection{Persian}
The Sharif Emotional Speech Database (ShEMO) was a database for the Persian language \cite{nezami2019shemo} and consisted of 3000 utterances from 87 native Persian speakers (31 females, 56 males). Six emotional categories were provided in the database: surprise, happiness, sadness, fear, anger, and neutral. In this experiment, we only consider happiness, sadness, anger, and neural. The dataset partition followed the same ratio as the ones above.

\subsubsection{German}
EMO-DB was an emotional speech database in the German language \cite{burkhardt2005database} and contained 10 German sentences from daily life communication produced by 10 actors (5 male and 5 female) in 7 emotions including anger, neutral, fear, joy, sadness, disgust, and boredom. Consistent with the data from the previous databases, four emotional categories (joy (happy), sadness, anger, and neutral) were selected for use in the experiment. Following the practice in \cite{kim2017towards}, audios from actors \#15 and \# 16 were used as test partition and validation partition respectively while audios from the rest of the actors were used as the training partition.

\begin{table*}[h!]
	\caption{UA and WA for unsupervised cross-lingual results.}
	\label{tab:allperformance}
	\centering
	\begin{tabular}{lcccccccc}
		\toprule
		\multirow{2}{*}{Method} & \multicolumn{2}{c}{Urdu} & \multicolumn{2}{c}{Ekropus} & \multicolumn{2}{c}{ShEMO} & \multicolumn{2}{c}{EMO-DB} \\ \cline{2-9} 
		& UA          & WA         & UA           & WA           & UA          & WA          & UA           & WA          \\ \hline
		SNSA-F                   & 36.36       & 36.54      & 22.17        & 23.12        & 22.31       & 20.24       & 31.25        & 41.18       \\
		SNSA-wo-SL                    & {\bf 56.63}       & {\bf 59.09}      & 27.30        & 28.31        & 27.70       & 35.01       & 43.75        & 47.06       \\
		SNSA-wo-HL                    & 48.12       & 47.72      & 24.21        & 24.67        & 28.18       & 28.87       & 46.02        & 50.00       \\ \hline
		NNPM                  & 54.55       & 51.31      & {\bf 28.34}        & {\bf 35.62}        & {\bf 28.19}       & {\bf 35.51}       & {\bf 50.57}        & {\bf 55.88}       \\ \bottomrule
	\end{tabular}
\end{table*}

\subsection{Spectrogram Extraction}

Firstly, the utterance duration is unified to 7.5 seconds by padding zeros for a short duration short than the unified duration and cropping along the time axis for long utterances which longer than the unified duration. Then, a Hanning window with a length of 400 is applied to the audio signals. The sampling rate is set at 16000Hz. For every frame, a short-term Fourier transform (STFT) of length 512 with hop length 160 is computed. Finally, Mel-scale is used to mimic the non-linear human ear perception of sound. We have also tried several different lengths of sampling window but the results had no clear differences.

\subsection{Experimental Setup}

The proposed model is trained in an end-to-end manner using the PyTorch toolkit \cite{paszke2019pytorch}. The IEMOCAP database is used for the source domain corpus and the rest databases are used as the target domain corpora. The model is trained for 50 epochs. In each iteration during the target domain training, the batch size is set to 32 and the model is optimized through the total loss by the Adam optimizer \cite{kingma2014adam} with a learning rate of $10^{-4}$, a decay rate of $5^{-5}$. A dropout is applied after every BLSTM layer with a 0.5 dropout probability. The dimensions of each SNSA module for the source and target domain setting are the same as \cite{li2019improved}. The similarity score threshold $\gamma$ is set to 0.9 and its influence will be studied in the ablation study of the experimental results. The memory updating rate $\beta$ starts from 0 and grows linearly to 0.4 through the 50 epochs of training. The hard negative sample ratio $\lambda$ is set to $0.01$. The parameters of the first two convolution modules of SNSA are frozen during training, and the influence of this setup will be studied in the ensuing ablation study of the experimental results.

The proposed NNPM model is compared against the following baselines:
\begin{itemize}
	\item {\bf SNSA-F:} the SNSA is trained on the source domain corpus. The parameters are frozen in order to save the source domain knowledge, which is then adapt to the target domain corpus directly by reusing the parameters of the source domain and a four emotional classifiers of the source domain. We refer to this model as the Frozen SNSA (SNSA-F).
	\item {\bf SNSA-wo-SL:} this model consists of SNSA and external memory with a pseudo-multilabeling process. The loss only contains the target domain loss $\mathcal{L}_\mathrm{hard}^t$ and no source domain loss. We refer to this model as the SNSA without Source domain Loss (SNSA-wo-SL).
	\item {\bf SNSA-wo-HL:} the setting of this baseline is similar to the SNSA-wo-SL, but the loss consists of both source domain loss and target domain loss without hard negative mining. We refer to this model as the SNSA without Hard mining Loss (SNSA-wo-HL).
	
\end{itemize}

\subsection{Experimental Results}

Both weighted accuracy (WA) and unweighted accuracy (UA) are measured as the evaluation criteria for evaluating the performances of the proposed and the baseline models. Our NNPM achieves 52.0 \% UA that training with IEMOCAP on the source domain and it contains 0.46 million parameters.

Table \ref{tab:allperformance} summarizes the UAs and WAs of the proposed NNPM and the three baseline models for the four low-resource corpora. Out of the four models examined in the experiment, the SNSA-F has the lowest performance across all corpora, suggesting that the source domain knowledge inside the SNSA-F alone is not enough for direct adaptation to the target domain. In comparison, the SNSA-wo-SL outperforms the SNSA-F by an absolute UA improvement of 20.27\% for Urdu, and 12.5\% for EMO-DB. This is likely due to the assignment of multilabel for target domain data by computing the similarities between features in the external memory and the target domain, which may have an improvement in the reliability of the target domain labels. But its performance is worse than NNPM. Since the source domain loss can guarantee the accuracy of the source domain module. The SNSA-wo-HL performance is worse than NNPM, which is because a large number of negative samples harm the model performance. The NNPM achieves the best overall performance, outperforming the SNSA-F by an absolute WA improvement of 17.77\% for Urdu, 12.50\% for Ekropus, 15.27\% for ShEMO, and 14.70\% for EMO-DB. These performance gains demonstrate the effectiveness of hard negative mining compares with SNSA-wo-SL. However, the performance of the NNPM is worse for Urdu compared to SNSA-wo-SL, which we speculate is due to a large difference in distribution between IEMOCAP and Urdu so that the source domain loss in NNPM may actually hurt the performance of target domain training. The performance for Ekropus and ShEMO under SNSA-F baseline is below 25\% due to a great domain mismatch between the English and Estonian or Persian. In addition, the backbone contains only knowledge of the source domain if the Siamese parameters are frozen without any fine-tuning for the target domain. The worse performance from Ekropus under SNSA-wo-HL is due to the greater unbalance in data distribution with almost 46\% of data being anger and only 1.7\% being joy. In comparison, the distribution of IEMOCAP is balanced so that it achieved 52.0\% UA on the source domain.

\begin{table}[h!]
	\centering
	\caption{Comparison UA with other state-of-the-art methods on the EMO-DB corpus. The DANN method \cite{abdelwahab2018domain} and AL method \cite{kim2017towards} results are taken from the re-implementation in paper \cite{ahn2021cross}. Therefore, we cite those results from \cite{ahn2021cross} directly.}
	\label{tab:emodbcompare}
	\begin{tabular}{lc}
		\toprule
		Method    & EMO-DB \\ \hline
		FLUDA \cite{ahn2021cross}   & 34.9   \\
		DANN \cite{ahn2021cross}     & 28.5   \\
		AL \cite{ahn2021cross} & 42.5   \\ \hline
		Ours (NNPM)     & {\bf 50.6}  \\ \bottomrule
	\end{tabular}
\end{table}
In addition to the experiment above, we also compare the proposed NNPM with three recent state-of-the-art methods, FLUDA \cite{ahn2021cross}, DANN \cite{abdelwahab2018domain} and AL \cite{kim2017towards}, for the EMO-DB database and the results are summarized in Table \ref{tab:emodbcompare}. The FLUDA is an unsupervised cross-corpus SER model based on few-shot learning \cite{ahn2021cross}. The NNPM largely outperforms this few-shot based method by a relative UA increase of 45.5\%. The domain adversarial neural network (DANN) and the aggregated multi-task learning (AL) are not publicly available, we do not implement these systems but instead cite these results of their performance on EMO-DB from re-implementations in \cite{ahn2021cross}. In comparisons, the NNPM outperforms adversarial-based method, that is DANN, by a ralative UA improvement of 77.5\% and outperformes multi-task learning based method, that is AL, with 19.1\%. The results show that the NNPM outperforms recent state-of-the-art methods. 

Finally, ablation studies are conducted to examine the influence of the modules of and hyperparameters in the NNPM on the system performance.
\begin{table}[h!]
	\centering
	\caption{Ablation study on freezing the first N convolutional layers of SNSA on the EMO-DB corpus.}
	\label{tab:freeze}
	\begin{tabular}{ccc}
		\toprule
		freeze N & UA & WA \\ \hline
		0 & 43.75 & 47.06 \\
		1 & 46.02 & 50.00 \\
		2 & {\bf 50.57} & {\bf 55.88} \\ 
		\toprule
	\end{tabular}
\end{table}

{\bf Influence of freezing the first N convolutional layers for SNSA.} Freezing a different number of the first N convolutional layers of SNSA and fine-tune the rest has different effects on the NNPM performance (see Table \ref{tab:freeze}). In specifies, freezing the first two layers layer and fine-tuning the rest gives the best performance and has a relative UA improvement of 9.9\% compared to freezing the first convolutional layer of SNSA and a relative UA improvement of 15.6\% compare to not freezing any convolutional layer on the EMO-DB corpus. Layer freezing can preserve knowledge from the source domain which can be reused for the target domain during the transfer learning. Because the first two layers usually contain more knowledge from the source domain, it is reasonable that freezing the first two layers in this experiment leads to the best model performance compared to freezing only the first layer or no layer freezing at all.

\begin{figure}[h!]
	\includegraphics[width=8cm]{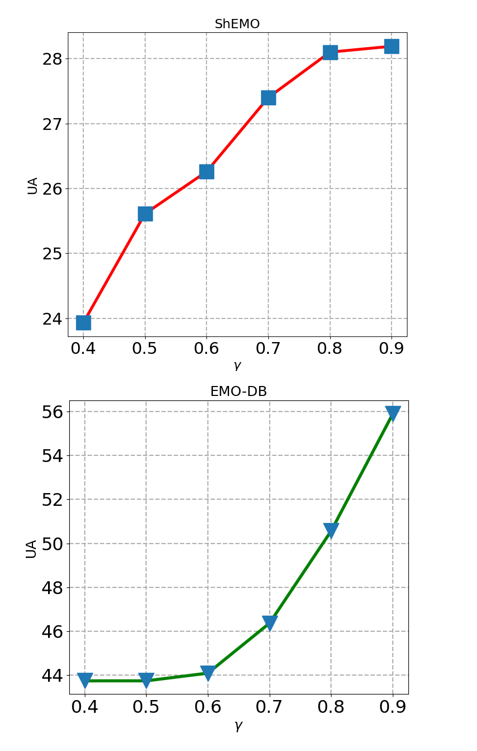}
	\caption{Ablation study of similarity score threshold $\gamma$ on ShEMO and EMO-DB corpora. }
	\label{fig:gamma}
\end{figure}
{\bf The similarity score threshold $\gamma$.} Hyper-parameter $\gamma$ controls the degree of similarity between features in the dynamic external memory and features from the target domain. The threshold is varied from 0.4 to 0.9 and evaluate on the ShEMO and EMO-DB corpora and the result is illustrated in Figure \ref{fig:gamma}. For the ShEMO corpus, the UA increases almost linearly from small to large $\gamma$, and the best UA is achieved when $\gamma =0.9$. For the EMO-DB corpus, a similar increasing trend is also found although the UA increases relatively slow at first and does not accelerate until $\gamma =0.6$. The best performance is also achieved when $\gamma =0.9$. This result shows that similarity between two features has a positive correlation with threshold $\gamma$ and that a large similarity score threshold may offer a more stable pseudo multilabel for the target domain data.

\section{Conclusions}

In this paper, we propose a novel unsupervised framework to improve the performance of cross-lingual SER with dynamic external memory and hard negative sample mining strategy. Experiments show that our NNPM exceeds baselines and other approaches across different low-resource corpora. 

\section{Acknowledgements}
This work was supported in part by National Key R\&D Program of China (2020YFC2004100), in part by National Natural Science Foundation of China \\ (NSFC U1736202, NSFC 61771461) and Shenzhen KQTD Project (No. KQTD20200820113106007).

% References should be produced using the bibtex program from suitable
% BiBTeX files (here: strings, refs, manuals). The IEEEbib.bst bibliography
% style file from IEEE produces unsorted bibliography list.
% -------------------------------------------------------------------------
\bibliographystyle{IEEEbib}
\bibliography{strings,refs}

\begin{thebibliography}{10}

\bibitem{el2011survey}
Moataz El~Ayadi, Mohamed~S Kamel, and Fakhri Karray,
\newblock ``Survey on speech emotion recognition: Features, classification
  schemes, and databases,''
\newblock {\em Pattern recognition}, vol. 44, no. 3, pp. 572--587, 2011.

\bibitem{cowie2001emotion}
Roddy Cowie, Ellen Douglas-Cowie, Nicolas Tsapatsoulis, George Votsis, Stefanos
  Kollias, Winfried Fellenz, and John~G Taylor,
\newblock ``Emotion recognition in human-computer interaction,''
\newblock {\em IEEE Signal processing magazine}, vol. 18, no. 1, pp. 32--80,
  2001.

\bibitem{albu2015neural}
Felix Albu, Daniela Hagiescu, Liviu Vladutu, and Mihaela-Alexandra Puica,
\newblock ``Neural network approaches for children’s emotion recognition in
  intelligent learning applications,''
\newblock in {\em EDULEARN15 7th Annu Int Conf Educ New Learn Technol
  Barcelona, Spain, 6th-8th}, 2015.

\bibitem{schuller2010cross}
Bjorn Schuller, Bogdan Vlasenko, Florian Eyben, Martin W{\"o}llmer, Andre
  Stuhlsatz, Andreas Wendemuth, and Gerhard Rigoll,
\newblock ``Cross-corpus acoustic emotion recognition: Variances and
  strategies,''
\newblock {\em IEEE Transactions on Affective Computing}, vol. 1, no. 2, pp.
  119--131, 2010.

\bibitem{zhou2019transferable}
Hao Zhou and Ke~Chen,
\newblock ``Transferable positive/negative speech emotion recognition via
  class-wise adversarial domain adaptation,''
\newblock in {\em ICASSP 2019-2019 IEEE International Conference on Acoustics,
  Speech and Signal Processing (ICASSP)}. IEEE, 2019, pp. 3732--3736.

\bibitem{latif2019unsupervised}
Siddique Latif, Junaid Qadir, and Muhammad Bilal,
\newblock ``Unsupervised adversarial domain adaptation for cross-lingual speech
  emotion recognition,''
\newblock in {\em 2019 8th International Conference on Affective Computing and
  Intelligent Interaction (ACII)}. IEEE, 2019, pp. 732--737.

\bibitem{ahn2021cross}
Youngdo Ahn, Sung~Joo Lee, and Jong~Won Shin,
\newblock ``Cross-corpus speech emotion recognition based on few-shot learning
  and domain adaptation,''
\newblock {\em IEEE Signal Processing Letters}, 2021.

\bibitem{abdelwahab2018domain}
Mohammed Abdelwahab and Carlos Busso,
\newblock ``Domain adversarial for acoustic emotion recognition,''
\newblock {\em IEEE/ACM Transactions on Audio, Speech, and Language
  Processing}, vol. 26, no. 12, pp. 2423--2435, 2018.

\bibitem{kim2017towards}
Jaebok Kim, Gwenn Englebienne, Khiet~P Truong, and Vanessa Evers,
\newblock ``Towards speech emotion recognition" in the wild" using aggregated
  corpora and deep multi-task learning,''
\newblock {\em arXiv preprint arXiv:1708.03920}, 2017.

\bibitem{goodfellow2016nips}
Ian Goodfellow,
\newblock ``Nips 2016 tutorial: Generative adversarial networks,''
\newblock {\em arXiv preprint arXiv:1701.00160}, 2016.

\bibitem{zhang2019limitations}
Huan Zhang, Hongge Chen, Zhao Song, Duane Boning, Inderjit~S Dhillon, and
  Cho-Jui Hsieh,
\newblock ``The limitations of adversarial training and the blind-spot
  attack,''
\newblock {\em arXiv preprint arXiv:1901.04684}, 2019.

\bibitem{ochal2020comparison}
Mateusz Ochal, Jose Vazquez, Yvan Petillot, and Sen Wang,
\newblock ``A comparison of few-shot learning methods for underwater optical
  and sonar image classification,''
\newblock {\em arXiv preprint arXiv:2005.04621}, 2020.

\bibitem{opitz1999popular}
David Opitz and Richard Maclin,
\newblock ``Popular ensemble methods: An empirical study,''
\newblock {\em Journal of artificial intelligence research}, vol. 11, pp.
  169--198, 1999.

\bibitem{li2019improved}
Yuanchao Li, Tianyu Zhao, and Tatsuya Kawahara,
\newblock ``Improved end-to-end speech emotion recognition using self attention
  mechanism and multitask learning.,''
\newblock in {\em Interspeech}, 2019, pp. 2803--2807.

\bibitem{hochreiter1997long}
Sepp Hochreiter and J{\"u}rgen Schmidhuber,
\newblock ``Long short-term memory,''
\newblock {\em Neural computation}, vol. 9, no. 8, pp. 1735--1780, 1997.

\bibitem{nair2010rectified}
Vinod Nair and Geoffrey~E Hinton,
\newblock ``Rectified linear units improve restricted boltzmann machines,''
\newblock in {\em Icml}, 2010.

\bibitem{grossberg1987competitive}
Stephen Grossberg,
\newblock ``Competitive learning: From interactive activation to adaptive
  resonance,''
\newblock {\em Cognitive science}, vol. 11, no. 1, pp. 23--63, 1987.

\bibitem{shrivastava2016training}
Abhinav Shrivastava, Abhinav Gupta, and Ross Girshick,
\newblock ``Training region-based object detectors with online hard example
  mining,''
\newblock in {\em Proceedings of the IEEE conference on computer vision and
  pattern recognition}, 2016, pp. 761--769.

\bibitem{busso2008iemocap}
Carlos Busso, Murtaza Bulut, Chi-Chun Lee, Abe Kazemzadeh, Emily Mower, Samuel
  Kim, Jeannette~N Chang, Sungbok Lee, and Shrikanth~S Narayanan,
\newblock ``Iemocap: Interactive emotional dyadic motion capture database,''
\newblock {\em Language resources and evaluation}, vol. 42, no. 4, pp.
  335--359, 2008.

\bibitem{latif2018cross}
Siddique Latif, Adnan Qayyum, Muhammad Usman, and Junaid Qadir,
\newblock ``Cross lingual speech emotion recognition: Urdu vs. western
  languages,''
\newblock in {\em 2018 International Conference on Frontiers of Information
  Technology (FIT)}. IEEE, 2018, pp. 88--93.

\bibitem{pedregosa2011scikit}
Fabian Pedregosa, Ga{\"e}l Varoquaux, Alexandre Gramfort, Vincent Michel,
  Bertrand Thirion, Olivier Grisel, Mathieu Blondel, Peter Prettenhofer, Ron
  Weiss, Vincent Dubourg, et~al.,
\newblock ``Scikit-learn: Machine learning in python,''
\newblock {\em the Journal of machine Learning research}, vol. 12, pp.
  2825--2830, 2011.

\bibitem{vcermakhlt}
Franti{\v{s}}ek {\v{C}}ermak, R{\=u}ta Marcinkevi{\v{c}}ien{\.e}, Erika
  Rimkut{\.e}, and Jolanta Zabarskait{\.e},
\newblock ``Hlt'2007: The estonian emotional speech corpus: Release,''
\newblock .

\bibitem{nezami2019shemo}
Omid~Mohamad Nezami, Paria~Jamshid Lou, and Mansoureh Karami,
\newblock ``Shemo: a large-scale validated database for persian speech emotion
  detection,''
\newblock {\em Language Resources and Evaluation}, vol. 53, no. 1, pp. 1--16,
  2019.

\bibitem{burkhardt2005database}
Felix Burkhardt, Astrid Paeschke, Miriam Rolfes, Walter~F Sendlmeier, and
  Benjamin Weiss,
\newblock ``A database of german emotional speech,''
\newblock in {\em Ninth European Conference on Speech Communication and
  Technology}, 2005.

\bibitem{paszke2019pytorch}
Adam Paszke, Sam Gross, Francisco Massa, Adam Lerer, James Bradbury, Gregory
  Chanan, Trevor Killeen, Zeming Lin, Natalia Gimelshein, Luca Antiga, et~al.,
\newblock ``Pytorch: An imperative style, high-performance deep learning
  library,''
\newblock {\em arXiv preprint arXiv:1912.01703}, 2019.

\bibitem{kingma2014adam}
Diederik~P Kingma and Jimmy Ba,
\newblock ``Adam: A method for stochastic optimization,''
\newblock {\em arXiv preprint arXiv:1412.6980}, 2014.

\end{thebibliography}

\end{document}